\documentclass[11pt]{article}
\usepackage{verbatim,url,enumerate,color,paralist}
\usepackage{amsmath,amsfonts}
\usepackage{nccbbb}
\usepackage{epsfig,amssymb,amstext,xspace,theorem}
\usepackage{algorithm}
\usepackage{algorithmicx,algpseudocode}
\usepackage{fullpage}

\newcommand{\qed}{\hspace*{\fill}$\Box$}
\newtheorem{theorem}{Theorem}[section]
\newtheorem{lemma}[theorem]{Lemma}

\newtheorem{definition}[theorem]{Definition}

\newtheorem{proposition}[theorem]{Proposition}
\newtheorem{claim}[theorem]{Claim}
\newtheorem{fact}[theorem]{Fact}

\newcommand{\IGNORE}[1]{}

\newcommand{\tree}{J}
\newcommand{\eventree}{\widehat{J}}
\newcommand{\cost}{\mbox{\rm cost}}
\newcommand{\lamfam}{\mathcal L}

\newcommand{\nodepartition}{\mathcal P}
\newcommand{\oddfamily}{\mathcal Q}
\newcommand{\evenfamily}{\mathcal R}
\newcommand{\oddactive}{{\mathcal C}}
\newcommand{\evenactive}{\widehat{\mathcal C}}
\newcommand{\isolatedactive}{\widehat{\mathcal I}}

\newcommand{\tspecial}{\widehat{t}}
\newcommand{\troot}{{t^{\star}}}

\newenvironment{proof}[1][Proof. ]{\noindent {\bf #1 }}{\qed}

\newcommand{\Xomit}[1]{}

\begin{document}

\title{Approximating Minimum-Cost Connected $T$-Joins}
\author{
Joseph Cheriyan\thanks{
	(jcheriyan@uwaterloo.ca)
	Dept.\ of Comb.\ \& Opt.,
	University of Waterloo, Waterloo, Ontario N2L3G1, Canada.
	}
\and
Zachary Friggstad\thanks{
        (zfriggstad@uwaterloo.ca)
	Dept.\ of Comb.\ \& Opt.,
        University of Waterloo, Waterloo, Ontario N2L3G1, Canada.
	}
\and
Zhihan Gao\thanks{
        (z9gao@uwaterloo.ca)
	Dept.\ of Comb.\ \& Opt.,
        University of Waterloo, Waterloo, Ontario N2L3G1, Canada.
	}
}

\date{\today}

\begin{titlepage}

\maketitle
\thispagestyle{empty}


\begin{abstract}
We design and analyse approximation algorithms for the \emph{minimum-cost
connected $T$-join} problem: given an undirected graph $G=(V,E)$
with nonnegative costs on the edges, and a set of nodes $T\subseteq{V}$,
find (if it exists) a spanning connected subgraph $H$ of minimum
cost such that every node in $T$ has odd degree and every node not
in $T$ has even degree; $H$ may have multiple copies of any edge
of $G$. Two well-known special cases are the TSP ($T=\emptyset$)
and the $s,t$~path TSP ($T=\{s,t\}$). Recently, An, Kleinberg, and
Shmoys [STOC 2012] improved on the long-standing $\frac{5}{3}$ approximation
guarantee for the latter problem and presented an algorithm based
on LP rounding that achieves an approximation guarantee of
$\frac{1+\sqrt{5}}{2}\approx1.61803$.

We show that the methods of An et al.\ extend to the minimum-cost
connected $T$-join problem.  They presented a new proof for a
$\frac53$ approximation guarantee for the $s,t$~path TSP; their
proof extends easily to the minimum-cost connected $T$-join problem.
Next, we improve on the approximation guarantee of $\frac53$ by
extending their LP-rounding algorithm to get an approximation
guarantee of
 $ \frac{13}{8}=1.625$ for all $|T|\geq 4$.
%
%

Finally, we focus on the prize-collecting version of the problem,
and present a primal-dual algorithm that is
``Lagrangian multiplier preserving'' and that achieves
an approximation guarantee of $3-\frac{2}{|T|-1}$ when $|T|\ge4$.
Our primal-dual algorithm is a generalization of the known
primal-dual 2-approximation for the prize-collecting $s,t$~path TSP.
Furthermore, we show that our
analysis is tight by presenting instances with $|T| \geq4$
such that the cost of the solution found by the algorithm is exactly
$3-\frac{2}{|T|-1}$ times the cost of the constructed dual solution.
\end{abstract}

\end{titlepage}


\section{Introduction}


The Traveling Salesman Problem (TSP) and its variants,
especially the $s,t$~path TSP,
are currently attracting substantial research interest.
We focus on a generalization that captures
the TSP and the $s,t$~path TSP.

Let $G = (V,E)$ be an undirected graph with nonnegative costs $c_e$
on the edges $e \in E$ and let $T$ be a subset of $V$.
A \emph{$T$-join} is a \emph{multiset} of edges $J$ of $G$
such that the set of nodes with odd degree
in the graph $H = (V,J)$ is precisely $T$,
that is, a node $v\in{V}$ has $\deg_J(v)$ odd if and only if $v\in{T}$,
\cite{edmonds:johnson,schrijver}.
A (spanning) \emph{connected $T$-join} is
a \emph{multiset} of edges $F$ of $G$
such that the graph $H = (V,F)$ is connected and
$T$ is the set of nodes with odd degree in $H$,
that is, a node $v\in{V}$ has $\deg_F(v)$ odd if and only if $v\in{T}$.
Clearly, we may (and we shall) assume that
$G$ is connected and that $|T|$ is even,
otherwise, no connected $T$-join exists;
moreover, we may assume that each edge of $G$ occurs with multiplicity
zero, one, or two in $H$,
otherwise, we may remove two copies of an edge from $H$ while
preserving the connected $T$-join property.
In the \emph{minimum-cost connected $T$-join} problem, the goal is
to find a connected $T$-join of minimum cost.
Two well-known special cases are
the TSP ($T=\emptyset$), and the $s$,$t$~path TSP ($T=\{s,t\}$).

By a \emph{metric graph} $G$ we mean a complete graph on $V(G)$
such that the edge costs satisfy the triangle inequality.
The \emph{metric completion} of a graph $G$ is given by the complete graph
on $V(G)$ with the cost of any edge $vw$
equal to the cost of a shortest $v,w$ path of $G$.
It can be seen that $G$ has a connected $T$-join of cost at most $\gamma$
if and only if the metric completion has a connected $T$-join of cost at most $\gamma$.
Thus, we may assume that the given graph $G$ is a metric graph.


Christofides presented an algorithm for the (metric) TSP that
achieves an approximation guarantee of $\frac{3}{2}$, \cite{christofides},
and this is the best result known for this problem.
Hoogeveen \cite{hoogeveen} extended the algorithm and its analysis
to the $s,t$~path TSP, and proved an approximation guarantee of $\frac53$.
Recently, An, Kleinberg, and Shmoys \cite{AKS-stoc12} improved on this
long-standing $\frac{5}{3}$ approximation guarantee
and presented an algorithm that achieves an approximation
guarantee of $\frac{1+\sqrt{5}}{2}\approx1.61803$.
To the best of our knowledge, there is {only} one previous result on
approximating min-cost connected $T$-joins:
Seb\H{o} and Vygen \cite{sebo:vygen} present
a very nice $\frac32$-approximation algorithm
for \emph{unweighted} graphs (each edge has unit cost);
in this context, we mention that the input graph cannot be
assumed to be a metric graph.
Seb\H{o} and Vygen \cite{sebo:vygen} were motivated in part
by previous advances on the special case of $T=\emptyset$
(namely, the graphic TSP) by
Oveis~Gharan, Saberi and Singh \cite{OSS-focs11},
M\"omke and Svensson \cite{momke:svensson}, and
Mucha \cite{mucha};
in fact, Seb\H{o} and Vygen \cite{sebo:vygen} achieve an
approximation guarantee of $\frac75=1.4$ for this special case.


All of our algorithms follow the plan of Christofides' algorithm:
first, compute an appropriate tree,
then, compute a $D$-join of minimum cost,
where $D$ denotes the set of nodes that have the ``wrong degree''
in the tree; finally, return the union of the tree and the $D$-join.
(Here, a $D$-join means a multiset of edges $E'$ such that
$D$ is the set of nodes of odd degree in $(V,E')$;
throughout the paper,
we use ``$T$'' and ``$T$-join'' as in the abstract, that is,
$T$ denotes a set of nodes specified in the input;
we use a symbol different from $T$ for a join with respect to
some auxiliary set of nodes.)


We show that the methods of An et al.\ extend to the minimum-cost
connected $T$-join problem.  They presented a new proof for a
$\frac53$ approximation guarantee for the $s,t$~path TSP; in
Section~\ref{sec:simple-approx}, we show that their proof extends
easily to the minimum-cost connected $T$-join problem.
More interestingly, in Section~\ref{sec:improve}, we generalize the
main result of An et al.\ to obtain an approximation guarantee of
$\frac{13}{8}=1.625<\frac53$ for $|T| \geq 4$. Our analysis uses some new methods
over that of An et al.\ and we elaborate in the next subsection.

Our second batch of results pertain to the
following prize-collecting version of the problem:
in addition to the graph $G=(V,E)$ and the edge costs $c$,
there is a nonnegative penalty $\pi(v)$ for each node $v\in
V\setminus{T}$; the goal is to find $I\subseteq V\setminus{T}$ and
a connected $T$-join $F$ of the graph $G\setminus{I}$ such that
$c(F) + \pi(I)$ is minimized.  The special
case of the prize-collecting TSP ($T=\emptyset$) has been extensively
studied for over 20 years, starting with Balas \cite{balas89}, and
an approximation guarantee of 1.91457 has been presented by Goemans
\cite{goemans09}; also see Archer et al.\ \cite{archer11}.  The
special case of the prize-collecting $s,t$~path TSP ($T=\{s,t\}$)
has also been studied, and An et al.\ \cite{AKS-stoc12} present
an approximation guarantee of 1.9535.

We focus on the general problem (prize-collecting connected $T$-join)
and present a primal-dual algorithm that achieves an approximation
guarantee of $3-\frac{2}{|T|-1}$ when $|T|\ge4$.  Our primal-dual
algorithm may be viewed as a generalization of the known primal-dual
2-approximation for the prize-collecting $s,t$~path TSP by Chaudhuri
et al.\ \cite{chaudhuri:godfrey:rao:talwar}, and we also match their
approximation guarantee of $2$ for $|T|=2$. Furthermore, we show
that our analysis is tight by presenting instances with $|T| \geq4$
such that the cost of the solution found by the algorithm is exactly
$3-\frac{2}{|T|-1}$ times the cost of the constructed dual solution.

In fact, the total penalty of the set of isolated nodes $I$ in the
solution found by our algorithm is at most~one times the penalty
incurred by the LP solution.  Thus, our algorithm has the ``Lagrangian
Multiplier Preserving'' property; this property is useful for the
design and analysis of approximation algorithms for cardinality-constrained
versions of problems.


Our algorithm and analysis follow Chaudhuri et al.\ 
\cite{chaudhuri:godfrey:rao:talwar}, and 
also we follow the well-known method of Goemans and Williamson
\cite{goemans:williamson95} for the prize-collecting
Steiner tree problem.
%
%
One key difference comes from the cost analysis for the $D$-join,
where $D$ denotes the set of nodes that have the wrong degree in
the tree computed by the algorithm. A simple analysis of the cost
of this $D$-join results in an approximation guarantee of
$4-O(|T|^{-1})$. To get the improved approximation guarantee, our
analysis has to go beyond the standard methods used for analysing
the approximation guarantee of primal-dual algorithms.

Most of our notation is standard, and follows Schrijver \cite{schrijver};
Section~\ref{sec:prelims} has a summary of our notation.

\subsection{ New Contributions on Min-Cost Connected $T$-Joins}
%

This subsection discusses the main points of difference between
our analysis and that of An et al.

Our algorithm and analysis follow that of An et al.\ at a high
level.  The algorithm solves an LP relaxation, and using the optimal
solution $x^*$ of the LP, it samples a random spanning tree $\tree$,
and then computes a min-cost $D$-join, where $D$ is the set of nodes
of the wrong degree in $\tree$.  The analysis hinges on
constructing a fractional $D$-join (a solution to an LP formulation
of the $D$-join problem) of low cost to ``fix'' the wrong-degree nodes in $\tree$.

We construct the fractional $D$-join as $y := \alpha \cdot \chi(\tree)
+ \beta \cdot x^* + z$ where
$\chi(\tree)$ is the 0-1 incidence vector for the edges
of $\tree$, $z$ is some ``correction'' vector
(described in Section~\ref{sec:vector}), and $\alpha$ and $\beta$ are
carefully chosen (scalar) values.
By the integrality of the $D$-join polyhedron, the cheapest $D$-join
has cost at most the cost of $y$.
By linearity of expectation, the expected cost of $y$ is less than or equal to
$\alpha + \beta$ times the cost of $x^*$ plus the expected cost of
$z$. It turns out that the correction vector $z$ is needed \emph{only}
for a special type of cut, the so-called $\tau$-narrow cuts: these
are given by $T$-odd sets $U$ such that $x^*(\delta(U))<1+\tau$.
When $|T|=2$, as in An et al.\ \cite{AKS-stoc12}, it turns out
that (the node sets of) the $\tau$-narrow cuts form a nested family
$U_1\subset U_2\subset \dots \subset U_i\subset \dots$.
This is no longer true for $|T|\ge4$, and hence,
the analysis of the correction vectors
by An et al.\ does not apply when $|T|\ge4$.

We prove that the $\tau$-narrow cuts form a laminar family
when $|T|\ge4$.
%
%
%
Moreover, in contrast with An et al., our analysis hinges on the
``partition inequalities'' that are satisfied by spanning trees and
fractional spanning trees such as $x^*$, namely, every partition
$\nodepartition = \{P_1, \ldots, P_k\}$ of the node set into nonempty
sets satisfies $x^*(\delta(P_1, \ldots, P_k)) \ge k-1$.  In our
application, we are given a subfamily of $\tau$-narrow cuts from
the laminar family of $\tau$-narrow cuts, and we have to obtain a
partition of the nodeset $V$ into \emph{nonempty} sets that correspond
to the given subfamily.  It is not clear that this holds for $\tau$
close to 1, but, we prove that it holds for $\tau \leq \frac{1}{2}$.


To complete the analysis, we have to fix $\alpha$, $\beta$ and $\tau$
subject to several constraints,
%
%
and we have to minimize the expected cost of the fractional $D$-join.
%
We choose $\tau=\frac12$, and this gives
$\alpha=\frac15$, $\beta=\frac25$;
moreover, we get a bound of $\frac{5}{8}\cost(x^*)$
on the expected cost of the fractional $D$-join,
and thus we get an approximation guarantee of $\frac{13}{8}=1.625$.
We have an example for $|T|=4$ showing that $\frac12$
is the optimal value for $\tau$ for our methods;
see Section~\ref{sec:tighteg}.

{
}


\section{Preliminaries \label{sec:prelims}}
We first establish some notation. Given a multiset of edges $F$,
we use $c(F)$ to denote the cost of $F$; thus, $c(F)=\sum_{e}\mu^F_e c_e$;
here, $\mu^F_e$ denotes the number of copies of the edge $e$ in $F$.

For any set of edges $F$ of $G$, we use $\chi(F)$ to denote the
zero-one incidence vector of $F$, thus, $\chi(F)\in\{0,1\}^{|E|}$, and
we use $V(F)$ to denote the set of incident nodes.
For any set of edges $F$ of $G$ and any subset of nodes $S$, we use
$F(S)$ to denote the set of edges of $F$ that have both endpoints in
$S$, and we use $\delta_F(S)$ to denote the set of edges of $F$ that
have exactly one endpoint in $S$.
We use the same notation for a multiset of edges.

For any set of nodes $S$, let $\overline{S}$ denote the complement
$V\setminus{S}$.
A set of nodes $S$ is called \emph{$T$-even} if $|S \cap T|$
is even, and it is called \emph{$T$-odd} if $|S \cap T|$ is odd.
Also, we say that a cut $\delta_F(S)$ is \emph{$T$-even}
(respectively, \emph{$T$-odd}) if $S$ is \emph{$T$-even}
(respectively, $S$ is \emph{$T$-odd}).
%

We say that two subsets of nodes $R$ and $S$ {\em cross} if
 $R \cap S,~ R \cup S,~ R \setminus S$ and $S \setminus R$
are all non-empty, proper subsets of $V$.
A family of subsets of $V$ is called \emph{laminar}
if no two of the subsets in the family cross. Equivalently,
a family of subsets of $V$ is laminar if
for any two subsets $R,S$ in the family,
either $R$ and $S$ are disjoint or one contains the other.

Let $\nodepartition = \{P_1, \ldots, P_k\}$ be a partition of the
nodes of $G$ into nonempty sets $P_1,\ldots,P_k$. Then
$\delta(\nodepartition)$ denotes the set of edges that have endpoints
in different sets in $\nodepartition$.

For ease of notation, we often identify a tree with its edge-set,
e.g., we may use $\tree\subseteq{E(G)}$ to denote a spanning tree.
Moreover, we may use relaxed notation for singleton sets, e.g.,
for a node $t$, we may use $V-t$ instead of $V\setminus\{t\}$.

We use the the next fact throughout the paper.
It relates the number of odd-degree nodes in a set $U\subseteq{V}$ and
the parity of the cut $\delta(U)$.

\begin{lemma} \label{lem:cutparity}
Let $G=(V,E)$ be a graph, and let $T\subseteq{V}$ have even size.
Let $F$ be a multiset of edges of $G$,
and let $D$ be the set of wrong-degree nodes w.r.t.\ $F$,
that is, $D$ consists of nodes $v\in{T}$ with $\deg_F(v)$ even
and nodes $v\in{V\setminus{T}}$ with $\deg_F(v)$ odd. Then,
for any $U\subseteq{V}$ we have
\begin{itemize}
\item[(i)]
$\displaystyle |\delta_F(U)| \equiv
	|U \cap D \cap \overline T| + |U \cap \overline D \cap T|
	\quad \pmod{2}$;
\item[(ii)]
moreover, if $U$ is both $T$-odd and $D$-odd,
then $|\delta_F(U)|$ is even.
\end{itemize}
\end{lemma}
\begin{proof}
First, we prove~(i). Summing over the degrees in $F$ of all nodes in $U$
we have the equation
\[
\sum_{v \in U \cap D \cap T} |\delta_{F}(v)| +
  \sum_{v \in U \cap D \cap \overline T} |\delta_{F}(v)| +
  \sum_{v \in U \cap \overline D \cap T} |\delta_{F}(v)| +
  \sum_{v \in U \cap \overline D \cap \overline T} |\delta_{F}(v)|
 =  2|F(U)| + |\delta_F(U)|, \]
since each edge in $\delta_F(U)$ is counted once and each edge
in $F(U)$ is counted twice. Now, the degree, in $F$, of each node in $U \cap D \cap T$ and
$U \cap \overline D \cap \overline T$ is even and
the degree of each node in $U \cap D \cap \overline T$ and
$U \cap \overline D \cap T$ is odd.
Then~(i) follows by reducing modulo~2.

Now, consider~(ii).
Since $U$ is both $T$-odd and $D$-odd,
it can be seen that $|U \cap D \cap \overline T|$ and
$|U \cap \overline D \cap T|$ have the same parity.
Then, by (i), $|\delta_F(U)|$ is even.
\end{proof}


\subsection{An LP Relaxation}

We will assume that $G$ is a metric graph for both the 5/3-approximation
and its improvement.  If $T = \emptyset$, then any solution $F$
forms an Eulerian graph $H = (V,F)$; then the standard argument of
following an Eulerian walk and shortcutting past repeated nodes
yields a Hamiltonian cycle of no greater cost.  Otherwise, if $T
\neq \emptyset$, then the next result shows that there is a
minimum-cost solution subgraph $H = (V,F)$ that is a spanning tree;
the proof follows by generalizing the notion of shortcutting an
Eulerian walk.

\begin{proposition} \label{prop:tree}
Let $G=(V,E)$ be a metric graph, and
let $T\subseteq V$ have even cardinality.
Assume that $T \neq \emptyset$.
Given a connected $T$-join $F$, we can efficiently find a spanning tree
of $G$ of cost $\le c(F)$ that is also a connected $T$-join.
\end{proposition}
\begin{proof}
Let $F$ be a connected $T$-join in $G$. Suppose that either $F$ has multiple
copies of an edge of $G$ or $F$ is not acyclic. Then we give a
procedure for finding a connected $T$-join of smaller size and no
greater cost.  This procedure can be repeated until we find a
connected $T$-join that is simple and has no cycles.

Let $C := v_1v_2, v_2v_3, \ldots, v_{k-1}v_k, v_kv_1$ be a cycle
in $F$ for $k \geq2$, where the case $k=2$ means we are considering two
copies of an edge $v_1v_2$ in $F$.
We first claim that there is another edge in $F$ apart from $v_1v_2,
\ldots, v_{k-1}v_k, v_kv_1$ that has at least one of these $v_i$ as its
end node.  If $V \setminus \{v_1, \ldots, v_k\} \neq \emptyset$ then
this is true because $F$ is a connected $T$-join. Otherwise, $v_i \in
T$ for some $1 \leq i \leq k$ since $T \neq \emptyset$.  But $v_i$ has
degree two using the edges in $C$ and odd degree in the $T$-join $F$,
so there is some edge in $F$ incident to $v_i$ that has not been
included in $C$.

Suppose $uv_i$ is an edge in $F$ that is not listed among the edges in
$C$. Remove $uv_i$ and $v_iv_{i+1}$ from $F$ and add $uv_{i+1}$ to $F$
(where we let $v_{i+1}$ denote $v_1$ if $i = k$).  If $u = v_{i+1}$,
then we simply remove $uv_i$ and $v_iv_{i+1}$ without adding any edges.
Denote the resulting multiset of edges by $F'$.  By the triangle
inequality, we have $c(F') \le c(F)$.

The parity of the degrees of the nodes does not change, so $F'$ is
still a $T$-join. Furthermore, we claim that the graph $H=(V,F')$ is
connected.
 To see this, observe that $H=(V,F')$ has a walk $W'$
between a pair of nodes $v,w$ if and only if $(V,F)$ has a walk $W$
between $v,w$, because any occurrence of $v_iv_{i+1}$ in $W$ could be
replaced by the sequence of edges given by $C \setminus
\{v_iv_{i+1}\}$, similarly, $uv_i$ could be replaced by $uv_{i+1},C
\setminus \{v_iv_{i+1}\}$, and any occurrence of $uv_{i+1}$ in $W'$
could be replaced by $uv_i,v_iv_{i+1}$.

This completes the proof: in a metric graph,
given a connected $T$-join that has cycles or multi-edges, we can
find a connected $T$-join of smaller size and no greater cost,
assuming $T\not=\emptyset$.
\IGNORE{
  Consider any $\emptyset \subsetneq S \subsetneq V$.  If $v_iv_{i+1}$
  crosses $S$, then there must be some other edge in the cycle $C$
  that crosses $S$ and this edge was not removed from $F$.  If $uv_i$
  crosses $S$ and $v_iv_{i+1}$ does not cross $S$, then $uv_{i+1}$
  also crosses $S$. Finally, if neither $v_iv_{i+1}$ nor $uv_i$ cross
  $S$ then there is another edge in $F$ that crossed $S$ and this
  edge remains in $F'$. In any case, some edge in $F'$ crosses $S$
  so $F'$ is a connected $T$-join.
}
\end{proof}

Let $F$ be a connected $T$-join and consider any $T$-even subset of
nodes $S$. Observe that $|\delta_F(S)|$ is even;
this follows by applying Lemma~\ref{lem:cutparity} to $F$
and noting that the set of wrong-degree nodes $D$ is empty.
%
This fact and Proposition \ref{prop:tree} lead to our linear
programming relaxation (L.P.1) for the minimum-cost connected $T$-join
problem.  The optimal value of (L.P.1) gives a lower bound on the
minimum cost of a connected $T$-join, because
there exists an optimal connected $T$-join whose
incidence vector satisfies all the constraints of (L.P.1).



$
\begin{array}{rrcll}
\hbox{({\bf L.P.1})}\quad  {\rm minimize}: & \sum_{e \in E} c_e x_e & & \\
{\rm subject~to}: & x(E(S)) & \leq & |S|-1 & \forall S \subsetneq V, |S| \geq 2 \\
& x(E(V)) & = & |V|-1 & \\
& x(\delta(S)) & \geq & 2 & \forall \emptyset \subsetneq S \subsetneq V, |S \cap T| {\rm ~even} \\
& x_e & \geq & 0 & \forall e \in E
\end{array}
$

The preceding discussion shows that the optimal value of  this linear
program is a lower bound for the optimal cost for the connected
$T$-join problem when $T \neq \emptyset$.  Using the ellipsoid method,
we can solve this linear program efficiently. The first two constraints
assert that a feasible solution $x$ must be in the spanning tree
polytope and these can be separated over efficiently (see
\cite{lau:ravi:singh}). The last constraints say that the total
$x$-value assigned to edges crossing any particular $T$-even cut should
be at least 2. An efficient separation oracle for these constraints
was developed by Barahona and Conforti \cite{barahona:conforti}.


Finally, we recall a linear programming formulation for the minimum
cost $T$-join problem, assuming nonnegative costs.
The extreme points of this LP are integral, see \cite{schrijver},
meaning that the optimal value of this LP is equal to
the minimum cost of a $T$-join.
We call any feasible solution to the following linear program
a {\em fractional $T$-join}.

$
\begin{array}{rrcll}
\hbox{({\bf L.P.2})}\quad  {\rm minimize} : & \sum_{e \in E} c_e x_e & & \\
{\rm subject~to} : & x(\delta(U)) & \geq & 1 & \forall U \subseteq V, |U \cap T| {\rm ~odd} \\
& x_e & \geq & 0 & \forall e \in E
\end{array}
$


\section{A $\frac53$-Approximation Algorithm \label{sec:simple-approx}}

Hoogeveen \cite{hoogeveen} showed that Christofides' 3/2-approximation
algorithm for the TSP (the case when $T=\emptyset$) extends to give
a 5/3-approximation algorithm for the $s,t$~path TSP (the case when
$T=\{s,t\}$).  Later, An, Kleinberg, and Shmoys (AKS) \cite{AKS-stoc12}
proved that the 5/3-approximation guarantee holds with respect to
(the optimal value of) an LP relaxation for the $s,t$~path TSP.


It turns out that Christofides' algorithm generalizes to give a
5/3-approximation algorithm for the min-cost connected $T$-join
problem; this is observed in \cite{sebo:vygen}.
The (generalized) algorithm first computes a minimum
spanning tree $\tree\subseteq{E(G)}$.  Then let $D$ denote the set
of ``wrong degree'' nodes in $\tree$. That is, $D$ consists of the
nodes in $T$ that have even degree in $\tree$ and the nodes in $V
\setminus T$ that have odd degree in $\tree$. Let $M\subseteq{E(G)}$
be a minimum-cost $D$-join.  Then the multiset $F = \tree\cup{M}$
($F$ has two copies of each edge in $\tree\cap{M}$) forms a connected
$T$-join.
Thus the algorithm is combinatorial and does not require solving
any linear programs.
%
%
The next result uses the method of An et al.\ to show that
the algorithm achieves an approximation guarantee of $5/3$ w.r.t.\
the optimal value of the LP relaxation (L.P.1);
we include the proof, since it serves as an introduction to
our improved approximation algorithm that is presented in the next section.

\begin{theorem}[An, Kleinberg, and Shmoys \cite{AKS-stoc12}]
Let $x^*$ be an optimal solution for the linear programming
relaxation of the connected $T$-join problem, (L.P.1), and let $OPT_{LP}$
denote the optimal value $\sum_{e\in{E}}c_ex^*_e$.
Then the solution $F$ computed by the algorithm
has cost $\le \frac{5}{3} OPT_{LP}$.
\end{theorem}
\begin{proof}
The first two constraints of the linear program ensure that any
feasible solution $x$ is contained in the spanning tree polytope
of $G$, that is, $x$ is a convex combination of zero-one incidence
vectors of spanning trees of $G$, \cite{schrijver}. Let $\tree$ be
a minimum spanning tree; then, we have $c(\tree)\le OPT_{LP}$.

Let $y := \frac{1}{3} \; \chi({\tree}) + \frac{1}{3} x^*$; we claim
that $y$ is a fractional $D$-join. By the integrality of the $D$-join
polyhedron, this would show that the cost of the $D$-join $M$ is
$\le\frac{2}{3} OPT_{LP}$, and hence, the cost of $F$ is $\le\frac{5}{3}
OPT_{LP}$.

To see that $y$ is a fractional $D$-join, consider any set of nodes
$U$ that is $D$-odd.  If $U$ is also $T$-odd, then
Lemma~\ref{lem:cutparity} part~(ii) implies that $|\delta_\tree(U)|$
is even; moreover, $\tree$ is connected, hence, $|\delta_\tree(U)|$
has size $\geq 2$. Also, we have $x^*(\delta(U)) \geq 1$, hence,
we have $y(\delta(U)) \geq 1$.  Otherwise, if $U$ is $T$-even, then
$x^*(\delta(U)) \geq 2$ by the last constraints of the linear
program, and $\tree\cap\delta(U)$ has size $\geq 1$ since $\tree$
is connected.  Thus we have $y(\delta(U)) \geq 1$ in this case as
well.  Hence, $y(\delta(U)) \geq 1$ holds for every $D$-odd set
$U\subseteq{V}$, therefore, by (L.P.2), $y$ is a fractional $D$-join.
\end{proof}


\section{An Improved Approximation For $|T|\geq 4$ \label{sec:improve}}

In this section, we improve on the approximation guarantee of $5/3$
for the mincost connected $T$-join problem, by extending the
approximation algorithm and analysis by An et al.\ \cite{AKS-stoc12},
for the $s,t$~path TSP.  We assume $|T| \geq 4$, and we prove an
approximation guarantee of
 $\frac{13}{8}=1.625$. %
(We note that the analysis in \cite{AKS-stoc12} for the case $|T|=2$
applies also to the linear program (L.P.1); there is a minor
difference between the two LP relaxations since (L.P.1) does not
have degree constraints for the nodes; but, the degree constraints
in their LP are only required in their analysis  to show that their
LP solution is a convex combination of spanning trees.)

\IGNORE{
  (There is a minor difference between the two LP relaxations since
  we do not include degree constraints for the nodes, but we claim
  that their analysis can be followed in our case to prove that the
  integrality ratio of our LP relaxation is also
  $\le\frac{1 + \sqrt 5}{2}$ for the case $|T|=2$.)
}

\begin{theorem}
There is an algorithm (described in Section~\ref{sec:improved-algorithm})
that finds a connected $T$-join $F$ of cost at most
$\frac{13}{8}$ times the optimum
value of linear program (L.P.1).
\end{theorem}



\subsection{The Algorithm \label{sec:improved-algorithm}}

Let $x^*$ denote an optimal solution to the linear programming
relaxation for the minimum-cost connected $T$-join problem. The first
two constraints of the LP allow us to decompose $x^*$ as a convex
combination of incidence vectors of spanning trees. That is, there
exist spanning trees $\tree_1, \ldots, \tree_k$ and non-negative values
$\lambda_1, \ldots, \lambda_k$ summing to 1 such that $x^* =
\sum_{i=1}^k \lambda_i \; \chi(\tree_i)$. By Caratheodory's theorem, we
may assume $k \leq |E|+1$ and it is possible to find these spanning
trees in polynomial time, \cite{schrijver}. For each spanning tree
$\tree_i$, let $D_i$ denote the set of nodes that have the ``wrong''
degree in $\tree_i$, that is, $D_i$ consists of the nodes in $T$ that
have even degree in $\tree_i$ and the nodes in $V \setminus T$ that
have odd degree in $\tree_i$.  Let $M_i$ be a minimum cost $D_i$-join
and let $F_i$ be the multiset formed by the union of $M_i$ and
$\tree_i$.  Clearly, each $F_i$ is a connected $T$-join.  We output the
cheapest of these solutions.

It is easier to analyze a related randomized algorithm. Rather than
trying every tree $\tree_i$, our algorithm randomly selects a single
tree $\tree$ by choosing $\tree_i$ with probability $\lambda_i$. Since
the deterministic algorithm tries all such trees, the cost of the
solution found by the deterministic algorithm is at most the expected
cost of the solution found by this randomized algorithm. Let $D$ denote
the set of nodes of wrong degree in $\tree$, $M$ denote the
minimum-cost $D$-join, and $F$ denote the (multiset) union of $M$ and
$\tree$. The randomized algorithm returns $F$.

The expected cost of $F$ is the expected cost of $\tree$ plus the
expected cost of the $D$-join $M$.  The expected cost of the tree
$\tree$ is precisely the cost of $x^*$ since each edge $e$ has
probability precisely $x^*_e$ of appearing in $\tree$.  We will show
that the expected cost of $M$ is at most $\frac{5}{8}$ times the cost of $x^*$.


\subsection{Constructing the Fractional $D$-Join}

As in the proof of the $5/3$-approximation guarantee, we will
construct a fractional $D$-join. However, instead of using exactly
$\frac{1}{3}$ of $\chi(\tree)$ and $\frac{1}{3}$ of $x^*$, we will
construct the fractional $D$-join as $y := \alpha \cdot \chi(\tree)
+ \beta \cdot x^* + z$  where $x^*\in\bbbr^{|E|}$,
 $z$ is some ``correction'' vector in $\bbbr^{|E|}$ to
be described below, and $\alpha$ and $\beta$ are values which
will be specified shortly. Again, by the integrality of the
$T$-join polyhedron, the cost of $M$ will be at most the cost of
$y$. By linearity of expectation, the expected cost of $y$ will be
exactly $\alpha + \beta$ times the cost of $x^*$ plus the expected
cost of $z$.

The following lemma shows that for certain $\alpha$ and $\beta$, the
correction vector is not needed for many cuts. The proof is similar to
a result in \cite{AKS-stoc12}.

\begin{lemma} \label{lem:goodcut}
Suppose $\alpha + 2\beta \geq 1$. Then $(\alpha \cdot \chi(\tree) +
\beta\cdot x^*)(\delta(U)) \geq 1$ if $U$ is either
\\
(i) $T$-even, or
\\
(ii) $T$-odd and $D$-odd, with $x^*(\delta(U)) \geq \frac{1-2\alpha}{\beta}$.
\end{lemma}
\begin{proof}
First, suppose $U$ is $T$-even. Then $x^*(\delta(U)) \geq 2$ by the LP
constraints.  Since $\tree$ is connected, then $|\tree \cap \delta(U)|
\geq 1$. Therefore, we have $\alpha \cdot \chi(\tree) (\delta(U)) \geq
\alpha$ and $\beta \cdot x^* (\delta(U)) \geq 2\beta$; the sum of the
two terms is $\ge \alpha + 2\beta \geq 1$.

Now, consider part~(ii).
Suppose that $U$ is $T$-odd with $x^*(\delta(U)) \geq
\frac{1-2\alpha}{\beta}$.  Since $U$ is both $T$-odd and $D$-odd,
Lemma~\ref{lem:cutparity} part~(ii) implies that $|\tree\cap\delta(U)|$
is even; moreover, $\tree$ is a spanning tree, hence, $\tree$ has
$\ge2$ edges in $\delta(U)$. Consequently, we have $\alpha \cdot
\chi(\tree)(\delta(U)) \ge2\alpha$, and moreover, $\beta \cdot
x^*(\delta(U)) \ge 1-2\alpha$ by the assumption on $x^*(\delta(U))$;
the lemma follows, since the sum of the two terms is $\ge1$.
\end{proof}

It will be convenient to fix a particular node $\tspecial \in T$.
Unless otherwise specified, when discussing a cut of the graph we
will take the set $S\subseteq{V}$ representing the cut
to be such that $\tspecial \not\in S$,
thus the cut will be denoted
$\delta(S), S\subseteq V-{\tspecial}$.
%
%
As $T$-odd cuts of the graph that have small
$x^*$ capacity will be used frequently in our analysis, we employ the
following definition.

\begin{definition}
Let $\tau \ge0$.
A $T$-odd subset of nodes $S$ is called \emph{$\tau$-narrow}
if $x^*(\delta(S)) < 1 + \tau$.
\end{definition}

Using this definition, Lemma \ref{lem:goodcut} says that if $\alpha +
2\beta \geq 1$ with both $\alpha, \beta \geq 0$, then the vector
$\alpha \cdot \chi(\tree) + \beta \cdot x^*$ satisfies all constraints
defining the $D$-join polyhedron except, perhaps, the constraints
corresponding to $T$-odd, $\tau$-narrow cuts for $\tau \geq
\frac{1-2\alpha}{\beta} - 1$.

An et al.\ in \cite{AKS-stoc12}, proved that if $R$ and $S$ are
distinct $\tau$-narrow, $T$-odd cuts then either $S \subset R$ or $R
\subset S$.  A generalization of this result to connected $T$-joins is
the following.

\begin{lemma} \label{lem:laminar}
If $\tau \leq 1$ and $R$ and $S$ are distinct $\tau$-narrow
cuts, then $R$ and $S$ do not cross.
\end{lemma}
\begin{proof}
Assume, for the sake of contradiction, that $R$ and $S$ cross. There
are two cases to consider, depending on the cardinality of $R \cap S
\cap T$.  If $R \cap S \cap T$ is odd, then $R \setminus S$ and $S
\setminus R$ are nonempty, proper subsets of $V$ that have even
intersection with $T$.  But then we have
  \[2 + 2\tau > x^*(\delta(R)) + x^*(\delta(S)) \geq
    x^*(\delta(R \setminus S)) + x^*(\delta(S \setminus R)) \geq 2+2,\]
where the last inequality follows from the LP constraints applied to
the $T$-even sets $R \setminus S$ and $S \setminus R$. However, this
contradicts $\tau \leq 1$.

If, on the other hand, $R \cap S \cap T$ is even, then $R \cap S$ and
$R \cup S$ are nonempty, proper subsets of $V$ that have even
intersection with $T$. A similar contradiction can be reached in this
case using the inequality $x^*(\delta(R)) + x^*(\delta(S)) \geq
x^*(\delta(R \cap S)) + x^*(\delta(R \cup S)),$ where we have
$\emptyset\neq R\cap{S}, R\cup{S}\neq{V}$ because $R,S$ cross.
\end{proof}

Another way to state Lemma~\ref{lem:laminar} is that the $\tau$-narrow,
$T$-odd cuts of the graph form a laminar family $\lamfam$ of
nonempty subsets of $V \setminus \{\tspecial\}$.

The correction vector $z$ that we add to $\alpha \cdot \chi(\tree) +
\beta \cdot x^*$ for the $T$-odd, $\tau$-narrow cuts can be constructed
from the following lemma.  The main difference from the analogous
result in \cite{AKS-stoc12} is that we require a further restriction on
the size of $\tau$.

\begin{lemma} \label{lem:correction}
Let $\lamfam = \{U_i\}$ be the laminar family of $T$-odd, $\tau$-narrow cuts.
For $\tau \leq \frac{1}{2}$
there exists vectors $f^{U}\in\bbbr^{|E|}$,
one for each cut $U_i \in \lamfam$,
such that the following three conditions hold.
\begin{enumerate}
\item For each $U \in \lamfam$, $f^{U} \geq 0$
\item $\sum_{U \in \lamfam} f^U \leq x^*$
\item For each $U \in \lamfam$, $f^U(\delta(U)) \geq 1$
\end{enumerate}
\end{lemma}

The proof of this lemma is deferred to the next section. Assuming this
lemma, we will now show how to complete the analysis of the algorithm.
We now fix $\tau$ to be $\frac{1}{2}$. We also set $\alpha :=
\frac{1}{5}$ and $\beta := \frac{2}{5}$.  For these choices of parameters, we have $\alpha + 2\beta \geq
1$ and $\tau = \frac{1-2\alpha}{\beta} - 1$.

We construct the correction vector $z$ by including an appropriate
multiple of $f^U$ for each $D$-odd cut $U \in \lamfam$. Formally,
  \[ z = \sum_{\substack{U \in \lamfam\\|U\cap D| {\rm~odd} }}
    (1 - 2\alpha - \beta x^*(\delta(U))) \cdot f^U. \]
%
%
Since $x^*(\delta(U)) < 1 + \tau$ and $\tau = \frac{1-2\alpha}{\beta} -
1$, we have $1 - 2\alpha - \beta x^*(\delta(U)) \geq 0$ for each $U \in
\lamfam$ which shows $z \geq 0$. From this, Lemma \ref{lem:goodcut}
shows that $y(\delta(U)) \geq 1$ for each $D$-odd, $T$-even cut $U$ and
each $D$-odd, $T$-odd cut $U$ that is not $\tau$-narrow. Finally, if
$U$ is $D$-odd, $T$-odd and $\tau$-narrow (so $U \in \lamfam$), then
$f^U(\delta(U)) \geq 1$ so $y(\delta(U)) \geq 2\alpha + \beta
x^*(\delta(U)) + (1 - 2\alpha - \beta x^*(\delta(U))) = 1$.  Thus, we
have proved the next result.

\begin{lemma}
The vector $y$ is a fractional $D$-join.
\end{lemma}

We conclude the analysis by bounding the expected cost of $y$.
The next result states that the probability that a $T$-odd cut $U$ is
also $D$-odd is $\leq x^*(\delta(U))-1$; this is an immediate extension
of a similar statement in \cite{AKS-stoc12}.
%

\begin{fact} \label{fact:oddprobability}
Let $U$ be a $T$-odd set.
Suppose that $\tree$ is a random spanning tree
(obtained from $x^*$ by choosing $\tree_i$ with probability $\lambda_i$).
Then
$\displaystyle {\mathbf {Pr}}[|D \cap U| {\rm ~is~odd}] \leq x^*(\delta(U))-1$.
\end{fact}

Therefore,
\[
{\mathbf E}[\cost(y)] = (\alpha + \beta) \; \cost(x^*) +
	\sum_{U \in \lamfam} (1 - 2\alpha - \beta x^*(U))\cdot
	{\mathbf {Pr}}[|D \cap U| {\rm ~is~odd}] \cdot \cost(f^U).
\]
Now, for each $U \in \lamfam$ we can bound $(1 - 2\alpha - \beta
x^*(\delta(U)))\cdot{\mathbf {Pr}}[|D \cap U| {\rm ~is~odd}]$ by $(1 -
2\alpha - \beta x^*(\delta(U)))\cdot(x^*(\delta(U))-1)$. This is $\frac{-2x^*(\delta(U))^2 + 5x^*(\delta(U)) -3}{5}$.  For $x^*(\delta(U))$ bound between
1 and $\frac{3}{2}$, the maximum value of this function is
achieved at $x^*(\delta(U)) = \frac{5}{4}$
and its value is $\frac{1}{40}$.

So, the expected cost of $y$ is at most $(\alpha + \beta) \cdot
\cost(x^*) + \frac{1}{40} \cdot \sum_{U \in \lamfam} \; \cost(f^U)$.
Since $\sum_{U \in \lamfam} f^U \leq x^*$, we have the final bound on
the expected cost of $y$ being $(\alpha + \beta + \frac{1}{40}) \;
\cost(x^*)$.  Adding this to the expected cost of $\tree$, we have that
the expected cost of the connected $T$-join is at most
$\frac{13}{8} \cost(x^*)$. Note that this is strictly less than $\frac{5}{3}$.



\subsection{Tight Example for $\tau$}\label{sec:tighteg}

Here, we present an example for $|T|=4$ showing that $\frac12$
is the optimal value for $\tau$ for our methods.

Let $G=(V,E)$ be the complete graph on four nodes $K_4$, and let $T=V$.
It can be seen that $x\in\bbbr^{|E|}$
with $x_e=\frac12,\;\forall e\in E$,
satisfies all the constraints of the LP relaxation (L.P.1).
Choose any one node to be $\tspecial$;
recall that for any cut $\delta(S)$ of the graph,
we assume that the set $S$ representing the cut
is a subset of ${V}-\tspecial$.
Suppose that we choose a value strictly greater than $\frac12$ for $\tau$.
Then we have four $T$-odd, $\tau$-narrow cuts, namely,
the cuts of the three singletons $S=\{v\},\;v\in{V}-\tspecial$,
and the cut of $S={V}-\tspecial$;
each of these cuts $\delta(S)$ has $x(\delta(S))=\frac32<1+\tau$.
Clearly, Lemma~\ref{lem:correction} does not apply, because
the sum of $f^S(\delta(S))$ over the four $\tau$-narrow cuts
has to be $\ge4$, but we have $x(E)=3$,
hence, part~2 of Lemma~\ref{lem:correction} cannot hold.
On the other hand, the lemma holds for $\tau=\frac12$.


\subsection{The Correction Vector}\label{sec:vector}

We complete the analysis by proving Lemma~\ref{lem:correction}. As in
\cite{AKS-stoc12}, we set up a flow network and use the
max-flow/min-cut theorem to ensure a flow exists with the desired
properties. However, our analysis is complicated by the fact that the
sets in $\lamfam$ are laminar rather than simply nested.

Our argument on the existence of the desired flow uses the following
inequality for spanning trees.  For a connected graph $H$ and a
partition of $V(H)$ into $k$ non-empty sets, $\nodepartition =
\{P_1,\ldots,P_k\}$, the number of edges that have endpoints in
different sets in $\nodepartition$ is at least $k-1$, that is,
$|\delta_{E(H)}(\nodepartition)| \geq k-1$. Thus, as our vector $x^*$ is
a convex combination of (incidence vectors of) spanning trees, we
have $x^*(\delta(P_1, \ldots, P_k)) \geq k-1$, for any partition
$P_1, \ldots, P_k$ of $V(G)$ into nonempty sets.

Let $\lamfam'$ be a subfamily of $\lamfam$. For $U \in \lamfam'$, let
$g_{\lamfam'}(U)$ be the nodes in $U$ that are not found in any smaller
subset in $\lamfam'$. That is,
 \[ g_{\lamfam'}(U) = \{ v \in U : v \not\in W
	{\rm ~for~any~} W \in \lamfam' {\rm ~with~} W \subsetneq U\}.
 \]
The following result is the key to generalizing the argument in
\cite{AKS-stoc12} to our setting.

\begin{lemma} \label{lem:partition}
Suppose that $\tau \leq \frac{1}{2}$.
Let $\lamfam'$ be any subfamily of $\lamfam$.
The family of subsets
$\{g_{\lamfam'}(U) : U \in \lamfam'\} \cup
	\{V \setminus \bigcup_{W \in \lamfam'} W\}$
forms a partition of $V$, and each such subset is nonempty.
\end{lemma}

\begin{proof}
Each node $v$ in some subset in the family $\lamfam'$ is in
$g_{\lamfam'}(U)$ for some $U \in \lamfam'$ since $v$ is ``assigned''
to the smallest subset of $\lamfam'$ containing $v$.  All other nodes
appear in the set $V \setminus \bigcup_{W \in \lamfam'} W$.  By
construction, the sets are disjoint. It remains to prove that each of
the sets is nonempty.

Since $\tspecial$ is not in any subset in the family $\lamfam'$, it
must be that $V \setminus \bigcup_{W \in \lamfam'} W \neq \emptyset$.
For a set $U \in \lamfam'$, let $m_{\lamfam'}(U)$ be the maximal proper
subsets of $U$ in the subfamily $\lamfam'$. That is, $W \in \lamfam'$
is in $m_{\lamfam'}(U)$ if $W\subsetneq{U}$ and no other subset $W' \in
\lamfam'$ satisfies $W \subsetneq W' \subsetneq U$.  Note that
$g_{\lamfam'}(U) = U \setminus \bigcup_{W \in m_{\lamfam'}(U)} W$ and
the sets in $m_{\lamfam'}(U)$ are disjoint.

For the sake of contradiction, suppose that $g_{\lamfam'}(U) =
\emptyset$. Then $U$ is the disjoint union of the sets in
$m_{\lamfam'}(U)$. Since every set in $\lamfam'$ is $T$-odd, then
$|m_{\lamfam'}(U)|$ is also odd and we let $2k+1 = |m_{\lamfam'}(U)|$.
Note that $2k+1 \geq 3$ which implies $k\geq 1$.

Now we examine the quantity $X = x^*(\delta(U)) + \sum_{W \in
m_{\lamfam'}(U)} x^*(\delta(W))$. One the one hand, since $U$ and each
$W \in m_{\lamfam'}(U)$ are $\tau$-narrow cuts, then $X < (1 +
\tau) + (2k+1)(1 + \tau) = (2k+2)(1+\tau)$.  On the other hand, we consider the partition
$\nodepartition=\{W: W\in m_{\lamfam'}(U)\} \cup \{V \setminus U\}$ of $V$. We claim
that $2x^*(\delta(\nodepartition))\leq X$. To
see this, notice that any edge $e$ with ends in $V \setminus U$ and $W_0$ for some $W_0\in m_{\lamfam'}(U)$ is counted twice in $X$.
(Once for $\delta(U)$ and once for $\delta(W_0)$.) Similarly, for any edge $e$ with ends in different subsets $W_0, W_1$ in $m_{\lamfam'}(U)$ is
also counted twice.(Once for $\delta(W_0)$ and once for $\delta(W_1)$.) By the partition inequality, we have
$2(2k+1)\leq 2x^*(\delta(\nodepartition))\leq X<(2k+2)(1+\tau)$. Thus, $2(2k+1)< (2k+2)(1+\tau)$ which implies $\tau>\frac{k}{k+1}\geq \frac{1}{2}$ since $k\geq 1$. This contradicts $\tau\leq \frac{1}{2}$.
\end{proof}

\begin{proof}[Proof of Lemma \ref{lem:correction}]
We now finish construction of the vectors $f^U, U \in \lamfam$ by
describing the flow network. Create a directed graph with 4 layers of
nodes, where the first layer has a single source node $v_s$ and the
last layer has a single sink node $v_t$. We have a node $v_U$ for each
$\tau$-narrow cut $U \in \lamfam$ in the second layer, and a node $v_e$
for each edge $e\in E(G)$ in the third layer.  For each $U \in
\lamfam$, there is an arc from $v_s$ to $v_U$ with capacity 1. For each
edge $e$ of $G$, there is an arc from $v_e$ to $v_t$ with capacity
$x^*_e$.  Finally, for each $U \in \lamfam$ and each $e \in \delta(U)$
we have an arc from $v_U$ to $v_e$ with capacity $\infty$.

We claim that there is a flow from $v_s$ to $v_t$ that saturates each
of the arcs originating from $v_s$; this is proved below. From such a
flow, we construct the vectors $f^U$ for $U \in \lamfam$ by setting
$f^U_e$ to be the amount of flow sent on the arc from $v_U$ to $v_e$
(where we use $f^U_e = 0$ if $e \not\in \delta(U)$). We have $f^U \geq
0$ and, by the capacities of the arcs entering $v_t$, $\sum_{U \in
\lamfam} f^U \leq x^*$. Finally, since each $U \in \lamfam$ has the arc
from $v_s$ to $v_U$ saturated by one unit of flow, we have
$f^U(\delta(U)) \geq 1$. Thus, the vectors $f^U, U \in \lamfam$
satisfy the requirements of Lemma \ref{lem:correction}.

We prove the existence of this flow by the max-flow/min-cut theorem.
Let $S$ be any cut with $v_s \in S, v_t \not\in S$. If $S$ contains
some node $v_U$ for $U \in \lamfam$ but not $v_e$ for some
$e\in\delta(U)$, then the capacity of $S$ is $\infty$. Otherwise, let
$\lamfam_S$ denote the subfamily of sets $U \in \lamfam$ such that the
node $v_U$ representing $U$ is in $S$.
%
%
Then the total capacity of the arcs leaving $S$ is at least
\[ |\lamfam| - |\lamfam_S| +
   \sum_{\substack{e \in \delta(U)\\{\rm for~some~}U \in \lamfam_S}}
   x^*_e. \]

Consider the collection of sets $\nodepartition_S := \{g_{\lamfam_S}(U),
U \in \lamfam_S\} \cup \{V \setminus \bigcup_{W \in \lamfam_S} W\}$.
From Lemma~\ref{lem:partition}, each set in $\nodepartition_S$ is
nonempty and the sets of $\nodepartition_S$ form a partition of $V$.

Next, we claim that $e \in \delta(\nodepartition_S)$ if and only if
$e\in\delta(U)$ for some $U\in\lamfam_S$. Consider an edge
$e\in\delta(\nodepartition_S)$. If one endpoint of $e$ is in $V \setminus
\bigcup_{W \in\lamfam_S} W$, then the other endpoint lies in
$g_{\lamfam_S}(U)$ where $U$ is the smallest set in $\lamfam_S$
containing this endpoint. But then $e \in \delta(U)$ because $e$ has
exactly one endpoint in $U$. Otherwise, $e = uv$ has $u \in
g_{\lamfam_S}(U)$ and $v \in g_{\lamfam_S}(W)$ for distinct sets $U, W
\in \lamfam_S$.  Suppose, without loss of generality, that either $U
\subsetneq W$ or $U \cap W = \emptyset$. Then by definition of
$g_{\lamfam_S}(W)$, we cannot have $v \in U$. Therefore,
$e\in\delta(U)$.

Conversely, if $e = uv \in \delta(U)$ for some $U \in \lamfam_S$ with,
say, $u \in U$, then $u$ lies in $g_{\lamfam_S}(W)$ where $W$ is the
smallest set in $\lamfam_S$ containing $u$. Since $W \subseteq U$ and
$v \not\in U$, then $v$ must lie in a different set in
$\nodepartition_S$. Thus, $e \in \delta(\nodepartition_S)$.

This shows
\[\sum_{\substack{e \in \delta(U)\\{\rm for~some~}U \in \lamfam_S}}
	x^*_e = x^*(\nodepartition_S) \geq |\lamfam_S|, \]
where the inequality holds since $|\nodepartition_S| = |\lamfam_S| + 1$.
Therefore, the capacity of the cut $S$ is at least $|\lamfam|$. Since
this holds for all $v_s$,$v_t$ cuts $S$, then the maximum flow is at
least $|\lamfam|$.  Finally, the cut $S = \{v_s\}$ has capacity
precisely $|\lamfam|$ so the maximum $v_s$,$v_t$ flow saturates all of
the arcs exiting $v_s$.
\end{proof}


\section{Prize-Collecting Connected $T$-Joins \label{sec:prizecollecting} }

We start with a linear programming relaxation of the prize-collecting
problem. For notational convenience, we define a large penalty for each
node in $T$.
We also designate an arbitrary node $\troot\in T$ as the \emph{root} node.
The LP has a variable $Z_X$ for each set $X\subseteq{V-\troot}$
such that $Z_X=1$ indicates that $X$ is the set of isolated nodes
of an optimal integral solution; moreover, we have a cut constraint
for each nonempty subset $S$ of $V-\troot$; the requirement (r.h.s.\
value) of a cut constraint is 1 or 2, depending on whether the set
$S$ is $T$-odd or $T$-even.

Let $\oddfamily$ denote the $T$-odd subsets of $V -\troot$ and
let $\evenfamily$ denote the non-empty, $T$-even subsets of $V - \troot$.
Our LP relaxation is stated below.

$
\begin{array}{rrcll}
\hbox{({\bf L.P.3})}\quad  {\rm minimize}: & \displaystyle{\sum_e} c_ex_e + \displaystyle{\sum_{X \subseteq V-\troot}} \pi(X)Z_X & & \\
{\rm subject~to}:
 & x(\delta(Q)) & \geq & 1 & \forall~ Q \in \oddfamily \\
 & x(\delta(R)) + \displaystyle{\sum_{X : X \supseteq R, X \subseteq V-\troot}} 2Z_X & \geq & 2 & \forall~ R \in \evenfamily \\
 & x, Z & \geq & 0
\end{array}
$

Consider any solution to the prize-collecting connected $T$-join problem. 
Let $I \subseteq V - \troot$ denote the set of isolated nodes
and let $F$ denote the connected $T$-join of $G\setminus{I}$;
thus this solution incurs a total cost of
$c(F)$ for the edges in $F$
plus $\pi(I)$ for the penalties of the nodes in $I$.
We define an integral solution to (L.P.3) by taking $Z_I = 1$,
$Z_S = 0$ for all other subsets $S \subseteq V - \troot$, and moreover,
for each edge $e$, we take $x_e$ to be
the number of copies of $e$ used in $F$.
By construction, the cost of this solution $(x,Z)$ is equal to $c(F)+\pi(I)$.

For every $Q \in \oddfamily$,
observe that at least one edge of $\delta(Q)$ is in $F$
(since $F$ connects the nodes in $T$);
this justifies the first set of constraints in the LP relaxation.
Now, focus on the second set of constraints in the LP relaxation, and
consider any one set $R \in \evenfamily$ and its constraint in (L.P.3).
If $R \subseteq I$, then the constraint is satisfied
due to the term $2Z_I$ (in the left-hand side of the constraint).
Otherwise, if $R \not\subseteq I$ then
at least one edge in $\delta(R)$ is in $F$
(since $F$ connects the nodes in $\{\tspecial\}\cup R\setminus{I}$);
moreover, by Lemma~\ref{lem:cutparity}, $|\delta_F(R)|$ is even,
so at least two edges of $\delta(R)$ are in $F$;
hence, the constraint is satisfied if $R \not\subseteq I$.
The above discussion is summarized by the next result.

\begin{fact}
The optimal value of (L.P.3) is at most the optimal cost of
a prize-collecting connected $T$-join.
\end{fact}

\IGNORE{
It is also true that for any prize-collecting connected $T$-join
solution there are always at least 2 edges in $\delta(R)$ for any
$R \in \evenfamily$ with $R \cap T \neq \emptyset$ so we could have
refined (L.P.3) by creating yet another set of constraints for these
sets $R$ that do not include the penalty variables.  However, to
avoid clutter in the LP and its dual, we do not treat these sets
$R$ separately; the high penalty on nodes in $T$ will ensure that
our primal-dual algorithm connects all nodes in $T - \troot$ to $\troot$.
}

\IGNORE{
 \begin{proof}
 Suppose that $I \subseteq V - \troot$ is the set of nodes not
 included in the optimum prize-collecting connected $T$-join solution
 $F$.  Define an integral solution to (L.P.3) with $Z_I = 1$ and
 $Z_R = 0$ for all other subsets $R \subseteq V - \troot$.  For each
 edge $e$, set $x_e$ to the number of copies of edge $e$ used in
 $F$.  By construction, the cost of this solution $(x,Z)$ is equal
 to the cost of the connected $T$-join $F$ plus the total penalty
 of nodes in $I$.

 Consider a set $Q \in \oddfamily$. Since $Q \cap T \neq \emptyset$
 and $\troot \not\in Q$, then $F$ contains an edge in $\delta(Q)$
 so $x(\delta(Q)) \geq 1$. Now consider some set $R \in \evenfamily$.
 If $R \subseteq I$, then $2Z_I$ is a term in the sum in the second
 set of constraints so the left-hand side of this constraint is at
 least 2. Otherwise, if $R \not\subseteq I$ then $F$ contains at
 least one edge in $\delta(R)$.  By Lemma \ref{lem:cutparity},
 $x(\delta(R))$ is even so it must be at least 2.
 \end{proof}
}

The dual of (L.P.3) has a variable $y_Q$ for
each primal-constraint of the first type, and a variable $y_R$ for each
primal-constraint of the second type; thus, each $T$-odd set
$Q\subseteq V-\troot$ has a dual variable $y_Q$, and each $T$-even set
$\emptyset \subsetneq R\subsetneq V-\troot$ has a dual variable $y_R$.



$
\begin{array}{rrcll}
\hbox{({\bf L.P.4})}\quad  {\rm maximize}: &\displaystyle{\sum_{Q \in \oddfamily}} y_Q + \displaystyle{\sum_{R \in \evenfamily}} 2y_R & & \\
{\rm subject~to}:
& \displaystyle{\sum_{S \in \oddfamily \cup \evenfamily : e \in \delta(S)}} y_S  & \leq & c_e & \forall~ e \in E \\
& \displaystyle{\sum_{\substack{R \subseteq X, R \in \evenfamily}}} 2y_R &  \leq & \pi(X) & \forall~ X \subseteq V-\troot\\
& y & \geq & 0 \\
\end{array}
$

Consider the dual LP and a feasible solution $y$; we call an edge $e$
\emph{tight} if the constraint for $e$ holds with equality, and we call
a set of nodes $X$ \emph{$\pi$-tight} if the constraint for $X$ holds
with equality.

\subsection{The Primal-Dual Algorithm}

The algorithm proceeds in phases. In each phase, a partition
$\nodepartition$ of $V(G)$ is maintained; some sets in this partition
are \emph{active} and some are \emph{inactive}.
Throughout, the set containing the root, $\troot$, is taken to be inactive.
The initial partition consists of singletons $\{v\}$ for every $v \in V$.
Each of the sets $\{v\}, v\in V-\troot,$  is designated as active.
We initialize $y_S := 0$ for every subset $S$ of $V$.
Let $F$ denote the set of edges chosen
during the growing phase of the algorithm; we initialize $F := \emptyset$.

Each phase proceeds as follows. We simultaneously raise $y_S$ for
every active set $S$ in the current partition at a uniform rate.
(Recall that sets containing $\troot$ have no dual variables.
Since the algorithm designates such sets as inactive, it never
uses dual variables of such sets.)
The phase ends when either (i)~an edge becomes tight or
(ii)~an active subset of nodes $S$ becomes $\pi$-tight.
If the former occurs, then we pick any edge $e=vw$ that becomes tight;
its endpoints $v$ and $w$ must be
in different components of the current partition;
we add $e$ to $F$, and we merge the
components in the current partition containing $v$ and $w$;
we call the resulting new component inactive if it contains the root,
otherwise, we call the new component active.
If the latter occurs, that is, if an active subset
$S\subseteq{V}$ in the partition becomes $\pi$-tight,
then $S$ becomes inactive.
The algorithm terminates when there are no remaining active sets.


Standard arguments show that the dual solution at the end of the
algorithm is feasible and that the set of edges $F$ chosen throughout
the algorithm is acyclic.
 We prune our solution $F$ in the usual way.
 Namely, we iteratively discard any edge $e$ such that
 there exists an inclusion-wise maximal set $X$ that was inactive
 at some point of the algorithm \emph{and} $\delta(X) = \{e\}$;
 moreover, after this stage of pruning, we discard all remaining
 edges that are not in the component of $\troot$.
Let $\tree$ denote the remaining subset of edges.
The subgraph that remains after discarding the isolated nodes is
a tree $\tree$ containing the root $\troot$.
Furthermore, since each node in $T$ has a large penalty,
then $\tree$ contains all nodes in $T$.

Finally, let $D\subseteq{V(\tree)}$ denote the set of nodes that have
the wrong degree in the tree $\tree$.
We compute a minimum-cost $D$-join $M$ and finally,
we output $\tree \cup M$ as a connected $T$-join on $V(\tree)$.
Let $I$ denote the set of nodes not included in $\tree$, thus
$I=V\setminus{V(\tree)}$.

\subsection{Analysis of the Primal-Dual Algorithm}

%
%
%
%

Our argument for bounding the cost of the tree $\tree$ and the
penalties of the nodes in $I$ is similar to known arguments.  A simple
way to bound the cost of the $D$-join $M$ would be to pair the nodes in
$D$ using edge-disjoint paths in $\tree$, so that adding $M$ to $\tree$
at most doubles the cost of the set of edges used.  However, we can improve
on this simple analysis of the cost of the $D$-join
by scrutinizing the analysis of the dual growing phase.
The following theorem summarizes the cost bounds.

\begin{theorem}
The penalty of the nodes in $I$ is exactly
$
\displaystyle{2\sum_{X \subseteq I} y_X},
$
the cost of the tree $\tree$ is
\[
{ \leq\left(2-\frac{1}{|T|-1}\right) \sum_{Q \in \oddfamily} y_Q +
	2\sum_{\substack{R \in \evenfamily, R \nsubseteq I}} y_R},
\]
and the cost of the $D$-join $M$ is
\[
{ \leq\left(1-\frac{1}{|T|-1}\right)\sum_{Q \in \oddfamily} y_Q +
	2\sum_{\substack{R \in \evenfamily, R \nsubseteq I}} y_R}.
\]
\end{theorem}

Let $\rho(|T|)$ denote the approximation guarantee of our
algorithm; below, we show that $\rho(|T|)={3-\frac{2}{|T|-1}}$
for $|T|\ge4$, and $\rho(2)=2$.
Before presenting the proof, we remark that this shows
$ \cost(\tree\cup{M}) + \rho(|T|)\cdot \pi(I)$
is at most $\rho(|T|)$ times the cost of the dual solution $y$.


\begin{proof}
The equation for the penalty is standard and follows by construction
since $I$ (being the union of the $\pi$-tight inactive components that were
pruned) is $\pi$-tight.  The analysis for the cost of $\tree$ is nearly
identical to Goemans and Williamson's analysis~\cite{goemans:williamson95}
and is included in Appendix~\ref{sec:dualgrow} for completeness.
One minor difference in our analysis
comes from the fact that there are at most $|T|$ components
that are $T$-odd at any point in the execution, and we
exploit this fact to derive an approximation guarantee that is tight on
some examples.

%

To bound the cost of the $D$-join $M$, we consider a possibly different
$D$-join $M'$ obtained by pairing the nodes in $D$ with edge-disjoint
paths in $\tree$.  Clearly $c(M) \leq c(M')$ so it suffices to bound
the cost of $M'$.  Let $\eventree$ be the subset of $\tree$ consisting
of edges $e$ such that $\tree \setminus \{e\}$ consists of two $D$-even
components.  Note that $M' \cap \eventree = \emptyset$ since, by parity
arguments, any $D$-join must have an even number of edges crossing any
$D$-even cut, and each edge of $\tree$ is used at most once in $M'$.
The next claim is the key to the improved cost analysis for minimal
$D$-joins.

\begin{claim} \label{claim:parity}
Let $Q$ be a $T$-odd component from any step in the execution of the
algorithm.  Then at least one of the edges in $\delta_\tree(Q)$ belongs
to $\eventree$.
That is, $\delta_{M'}(Q)$ is a proper subset of $\delta_\tree(Q)$.
\end{claim}
\begin{proof}[Proof of Claim]
We have two cases to consider, either $Q$ is $D$-odd or it is
$D$-even.  First, suppose $Q$ is $D$-odd.  Then, by
Lemma~\ref{lem:cutparity} part~(ii), $|\delta_\tree(Q)|$ is even.
Focus on $\tree\setminus\delta_{\tree}(Q)$ and observe that it has an
odd number of connected components, so at least one of them, say $S$,
must be $D$-even.  Thus, the edge in $\delta_\tree(Q)$ connecting
$Q$ to $S$ is in $\eventree$.

Similarly, if $Q$ is $D$-even, then $|\delta_\tree(Q)|$ is odd.
Then $\tree\setminus\delta_{\tree}(Q)$ has an even number of
connected components, hence, there is another connected component
that is $D$-even, call it $S$, $S\not=Q$.
Then, the edge between $Q$ and $S$ is in $\eventree$.
\end{proof}

Using this, we can bound the cost of $M'$ in the following way.
\begin{eqnarray*}
  \sum_{e \in M'} c_e & = & \sum_{e \in M'}
	\left( \sum_{\substack{Q \in \oddfamily\\e \in \delta(Q)}} y_Q +
	\sum_{\substack{R \in \evenfamily\\ R \nsubseteq I, e \in \delta(R)}} y_R\right)
	= \sum_{Q \in \oddfamily} |\delta_{M'}(Q)| y_Q +
	\sum_{\substack{R \in \evenfamily, R \nsubseteq I}} |\delta_{M'}(R)| y_R \\
	& \leq & \sum_{Q \in \oddfamily} (|\delta_\tree(Q)|-1) y_Q +
	\sum_{\substack{R \in \evenfamily, R \nsubseteq I}} |\delta_\tree(R)| y_R
	\leq \left(1-\frac{1}{|T|-1}\right) \sum_{Q \in \oddfamily} y_Q +
	2\sum_{\substack{R \in \evenfamily, R \nsubseteq I}} y_R
\end{eqnarray*}
The first inequality follows from the claim for the $T$-odd sets in
$\oddfamily$ and the simple fact that $\delta_{M'}(R) \subseteq
\delta_\tree(R)$ for $R \in \evenfamily$.  The second inequality follows
from our analysis of the cost of $\tree$ in Appendix~\ref{sec:dualgrow}.
\end{proof}

This completes the analysis of the primal-dual algorithm.  Our
algorithm and analysis are also valid in the case $|T|=2$, and it
can be seen that our approximation guarantee for $|T|=2$ is
$\rho(2)=2$. In fact, our algorithm in this case is essentially
identical to the 2-approximation for the prize-collecting $s,t$~path TSP
presented in \cite{chaudhuri:godfrey:rao:talwar}.


Our analysis is tight even up to lower-order terms when $|T| \geq 4$.
This is realized by a cycle on $T$, that is,
$G=(T,E)$ consists of an even-length cycle with at least 4 nodes.
Let $\troot\in T$ be a designated node and let the edges incident
to it have cost $\frac12$ while all other edges have cost one.
The dual growth phase grows $y_{\{v\}}$
to $1/2$ for every singleton $v\in T-\troot$.
The algorithm could find a tree of cost $|T|-\frac32$
(by picking all edges of $G$ except one of the two edges incident to $\troot$),
and then find a $D$-join of cost $\frac{|T|-2}{2}$.
Observe that the cost of the dual solution is $\frac{|T|-1}{2}$,
whereas the connected $T$-join constructed by the algorithm
has cost $\frac{3|T|-5}{2}$;
the ratio of these two quantities is exactly $3-\frac{2}{|T|-1}$.



\section{Conclusions}

We presented a $\frac{13}{8}=1.625$ approximation algorithm for the
mincost connected $T$-join problem whose analysis closely followed
the analysis of the $s,t$~path TSP algorithm in \cite{AKS-stoc12}.
Furthermore, we presented a $\max\{3-\frac{2}{|T|-1},\;2\}$-approximation
algorithm for a prize-collecting version of the problem;
this algorithm is based on the primal-dual method~\cite{goemans:williamson95}
and it is Lagrangian multiplier preserving.

Our algorithms in Sections~\ref{sec:improve} and~\ref{sec:prizecollecting}
are based on the LP relaxations (L.P.1) in Section~\ref{sec:prelims}
and (L.P.3) in Section~\ref{sec:prizecollecting}, respectively.
Unfortunately, we do not have tight bounds on the integrality ratios
of these LP relaxations. As far as we know, the best
lower bound on the integrality ratio of (L.P.1) is $\frac32$, and
this follows from an example for the $s,t$~path TSP in
\cite[Figure~1]{AKS-stoc12}.

\bigskip
\noindent
{\bf Acknowledgements}: We thank a number of colleagues for
useful discussions; in particular, we thank Jochen K\"{o}nemann
and Chaitanya Swamy.



\vfill\clearpage

\appendix
\section{Appendix: Analysis of the Dual Growing Phase \label{sec:dualgrow}}

We bound the cost of $\tree$ as follows.

\[
\sum_{e \in \tree} c_e =
\sum_{e \in \tree} \left( \sum_{\substack{Q \in \oddfamily\\ e \in \delta(Q)}} y_Q +
  \sum_{\substack{R \in \evenfamily\\R \nsubseteq I, e \in \delta(R)}} y_R\right)  \\
= \sum_{Q \in \oddfamily} |\delta_\tree(Q)| y_Q +
  \sum_{\substack{R \in \evenfamily\\R \nsubseteq I}} |\delta_\tree(R)| y_R \\
\]

The first equation holds because the edges in $\tree$ are tight. That
the inner sum over subsets $R \in \evenfamily$ can be restricted to
subsets $\nsubseteq I$ follows because no subset of nodes
contributing to the dual constraint for an edge $e \in \tree$ is
contained in $I$.
%
%
The second equation follows by rearranging the sums.

Now consider a step in the execution with corresponding partition
$\nodepartition$ of $V(G)$. Add the edges of $\tree$ to the graph
$(V,\emptyset)$, and then contract each of the sets $S$ belonging to
the partition $\nodepartition$.
The resulting graph is a tree plus some isolated nodes, because
each contracted set $S$ of $\nodepartition$ induces
a tree of $(V,F)$ and so the subgraph of $(V,\tree)$ induced by $S$
consists of a tree plus some isolated nodes,
see \cite{goemans:williamson95,williamson:shmoys}.
Let $\oddactive$ denote the $T$-odd active sets in $\nodepartition$,
let $\evenactive$ denote the $T$-even active sets in $\nodepartition$
which are not contained in $I$, and let
$\isolatedactive$ denote the inactive sets $B \in \nodepartition$ with
$\delta_{\tree}(B) \neq \emptyset$
($\nodepartition$ could contain inactive sets $B$ with
$\delta_{\tree}(B) = \emptyset$, but such sets are not relevant for the
arguments below).
%
%
We can identify these sets with nodes in the contracted graph. It can
be seen that each $B \in \isolatedactive$, except for one,
has degree at least 2 in this contracted graph by our pruning phase;
if a set in $\isolatedactive$ contains the root,
then its degree could be one, see \cite[Chapter~14.1]{williamson:shmoys}.
Notice also that $|\oddactive| \leq |T|-1$
because each $T$-odd active set must contain a node in $T-\troot$.
By counting degrees, we have
 \begin{eqnarray*}
	2|\oddactive| + 2|\evenactive| + 2|\isolatedactive| - 2 & = &
	\sum_{Q \in \oddactive} |\delta_{\tree}(Q)| +
	\sum_{R \in \evenactive} |\delta_{\tree}(R)| +
	\sum_{B \in \isolatedactive} |\delta_{\tree}(B)| \\
 & \geq & \sum_{Q \in \oddactive} |\delta_{\tree}(Q)| +
	\sum_{R \in \evenactive} |\delta_{\tree}(R)| + 2|\isolatedactive| - 1,
 \end{eqnarray*}
hence,
 \[
	\sum_{Q \in \oddactive} |\delta_{\tree}(Q)| +
	\sum_{R \in \evenactive} |\delta_{\tree}(R)| \leq
	2|\oddactive| + 2|\evenactive| - 1 \leq
	\left(2-\frac{1}{|T|-1}\right)\cdot|\oddactive| +
	2|\evenactive|,
 \]
where the last inequality holds because $|\oddactive| \leq |T|-1$.
Suppose that the dual variables of the active sets were raised by
$\Delta$ during this phase. Then
 \[
	\sum_{Q \in \oddactive} \Delta |\delta_\tree(Q)| +
	\sum_{R \in \evenactive} \Delta |\delta_\tree(R)| \leq
	\left(2-\frac{1}{|T|-1}\right) \Delta |\oddactive| + 2 \Delta |\evenactive|.
 \]
Since this holds over each phase of the primal-dual algorithm, then
by applying induction on the number of phases in the execution, we have
 \[
	\sum_{Q \in \oddfamily} |\delta_\tree(Q)| y_Q +
	\sum_{\substack{R \in \evenfamily, R \nsubseteq I}}
		|\delta_\tree(R)| y_R \leq
	\left(2-\frac{1}{|T|-1}\right) \sum_{Q \in \oddfamily} y_Q +
	2 \sum_{\substack{R \in \evenfamily, R \nsubseteq I}} y_R.
 \]
This proves the bound on the cost of $\tree$.

\end{document}